\documentclass[showpacs,preprintnumbers,amsmath,amssymb,10pt]{revtex4}

\usepackage{psfig}
\usepackage{graphicx}% Include figure files
\usepackage{dcolumn}% Align table columns on decimal point
\usepackage{bm}% bold math

\newcommand{\s}{\ensuremath{\psi(t,r)}}
\newcommand{\n}{\ensuremath{\nu(t,r)}}
\newcommand{\T}{\ensuremath{\theta}}
\newcommand{\pt}{\ensuremath{p_\theta}}
\newcommand{\e}{equation}

\begin{document}
\preprint{}
\title{What role pressures play to determine the final end-state of 
gravitational collapse?}
\author{Rituparno Goswami}
\email{goswami@tifr.res.in}
\author{Pankaj S Joshi}
\email{psj@tifr.res.in}
\affiliation{ Department of Astronomy and Astrophysics\\ Tata 
Institute of Fundamental Research\\ Homi Bhabha Road,
Mumbai 400 005, India}

\begin{abstract}
We examine here in what way the pressures affect the 
final fate of a continual gravitational collapse. It is shown that
the presence of a non-vanishing pressure gradient in the collapsing 
cloud can determine directly the epoch of formation of trapped surfaces 
and the apparent horizon, thus changing the causal structure in
the vicinity of singularity. 
\end{abstract}
\par


\maketitle

\noindent Can non-zero pressures within a collapsing matter 
cloud avoid the naked singularity forming as end-state of a continual 
gravitational collapse? There have been speculations over past 
decades that when the effects of pressure within a collapsing 
cloud are carefully taken into account, may be only a black hole 
will form as the end product of collapse (see e.g. [1], for an early
mention to role of pressure in collapse, and [2] and [3] for further
details and references). 
It is clear that in the final stages of an endless collapse
pressures will certainly become quite important.  
Though many 
gravitational collapse models are known currently with non-vanishing 
pressures present, which end up in the formation of a black hole 
or a naked singularity (see e.g. [4],[5] for a recent discussion), 
the actual role the pressures play to determine 
the end state of a continual collapse is however not clearly understood.

We analyze this issue here in some detail in order to bring out
the role pressures can play towards 
determining the end state 
of collapse. A spherically symmetric collapse model is considered 
where the initial density and radial pressure distributions are 
chosen to be homogeneous, but the tangential pressure is allowed to
have a non-vanishing gradient, in order to see in a transparent manner
the effects it can cause on the evolution of the cloud, and eventually
towards determining the final state of collapse. It 
turns out that this by itself can 
cause inhomogeneities in the density evolution as the collapse 
develops, thus deforming the trapped surface formation within the 
cloud. Hence it is seen that the presence of a non-vanishing pressure can 
create  
either of the naked singularity or a black hole as 
the final state for the cloud. Further, the pressure diverges 
along various families of non-spacelike curves terminating into 
the naked singularity. 
This is relevant because if a naked singularity developed but
if the pressures remained finite in the limit of approach to the
same, this may not be regarded as a physically interesting situation.

The spherically symmetric metric in a general form can be 
written as, 
\begin{equation}
ds^2=-e^{2\n}dt^2 + e^{2\s}dr^2 + R^2(t,r)d\Omega^2
\label{eq:metric}
\end{equation}
where $d\Omega^2$ is the line element on a two-sphere. Choosing 
the frame to be comoving, the stress-energy tensor for a general (type I)
matter field is given in a diagonal form [6],
\begin{equation}
T^t_t=-\rho;\; T^r_r= p_r;\; T^\T_\T=T^\phi_\phi= p_\T
\label{eq:setensor}
\end{equation}
The quantities $\rho$, $p_r$ and $p_\T$ are the density, radial 
and tangential pressures respectively. We take the matter field to 
satisfy the {\it weak energy condition}, that is, the energy density as
measured by 
any local observer be non-negative, and for any timelike vector 
$V^i$, we have,
\begin{equation}
T_{ik}V^iV^k\ge0
\end{equation}
which amounts to,
\begin{equation}
\rho\ge0;\; \rho+p_r\ge0;\; \rho+p_\T\ge0
\end{equation}
The initial data consists of values of three metric functions
and the density and pressures at the initial time $t=t_i$, 
in terms of the six arbitrary functions of the radial coordinate,
$\nu(t_i,r)=\nu_0(r), \psi(t_i,r)=\psi_0(r), R(t_i,r)=r,
\rho(t_i,r)=\rho_0(r), p_r(t_i,r)=p_{r0}(r), \pt(t_i,r)=p_{\T0}$,
where, using the scaling freedom for the radial co-ordinate $r$ 
we have chosen $R(t_i,r)=r$ at the initial epoch. The dynamic 
evolution of the initial data is then determined by the Einstein 
equations, which for the metric (\ref{eq:metric}) become ($8\pi G=c=1$),
\begin{eqnarray}
\rho=\frac{F'}{R^2R'}; && p_r=\frac{-\dot{F}}{R^2\dot{R}}
\label{eq:ein1}
\end{eqnarray}
\begin{equation}
\nu'=\frac{2(\pt-p_r)}{\rho+p_r}\frac{R'}{R}-\frac{p_r'}{\rho+p_r}
\label{eq:ein2}
\end{equation}
\begin{equation}
-2\dot{R}'+R'\frac{\dot{G}}{G}+\dot{R}\frac{H'}{H}=0
\label{eq:ein3}
\end{equation}
\begin{equation}
G-H=1-\frac{F}{R}
\label{eq:ein4}
\end{equation}
Here $F=F(t,r)$ is an arbitrary function, and in spherically symmetric 
spacetimes it has the interpretation of the mass function, with $F\ge0$. 
In order to preserve the regularity of the initial data, 
$F(t_i,0)=0$, i.e. the mass function should vanish at the center 
of the cloud. The functions $G$ and $H$ are defined as
$G(t,r)=e^{-2\psi}(R')^2$ and $H(t,r)=e^{-2\nu}(\dot{R})^2$.

All the initial data above are 
not mutually independent, as from equation (\ref{eq:ein2}) we 
find that the function $\nu_0(r)$ is determined in terms of
rest of the initial data.
Also, by rescaling of the radial coordinate $r$ we have reduced 
the number of independent initial data functions to four. We then 
have a total of five field equations with seven unknowns, $\rho$, 
$p_r$, $\pt$, $\psi$, $\nu$, $R$, and $F$, giving us the freedom 
of choice of two free functions. Selection of these functions, subject 
to the given initial data and weak energy condition, determines 
the matter distribution and metric of the space-time and thus 
leads to a particular collapse evolution of the initial data.

Consider, for the sake of clarity, the following choice of the 
allowed free functions, $F(t,r)$ and $\n$ (see also [7]),
\begin{equation}
F(t,r)=\frac{2}{3}r^3-\frac{1}{3}R^3
\label{eq:mass}
\end{equation}
and,
\begin{equation}
\n=c(t)+\nu_0(R)
\label{eq:nu}
\end{equation}
Also we write,
\begin{equation}
R(t,r)=rv(t,r)
\label{eq:R}
\end{equation}
where,
\begin{eqnarray}
v(t_i,r)=1; & v(t_s(r),r)=0; & \dot{v}<0
\label{eq:v}
\end{eqnarray}
Using equation (\ref{eq:mass}) in equation (\ref{eq:ein1}), we get,
\begin{eqnarray}
\rho=\frac{2}{v^2(v+rv')}-1; && p_r=1
\label{eq:rho} 
\end{eqnarray}

It is clear that $\rho(t_i,r)=\rho_0(r)=1$, i.e. at the initial 
epoch the density is homogeneous, and as $v\rightarrow 0$, $\rho\rightarrow
\infty$. Hence, at the singularity $\rho$ becomes infinite. Also, we 
note that the radial pressure remains constant throughout the 
collapse. However, this need not be the case for the tangential
pressure, which depends via the Einstein equation (8) 
on the choice of the function $\nu$. Then using 
\e (\ref{eq:rho}) we have
\begin{equation}
\nu_0(r)=\int_0^r\left(\frac{p_{\T0}-1}{r}\right)dr
\label{eq:nu01}
\end{equation}  
Assuming that pressure gradients vanish at the center of
the cloud, we can take the form of $\nu_0(r)$ as,
\begin{equation}
\nu_0(r)=r^2g(r)
\label{eq:nu0form}
\end{equation} 
where, $g(r)$ is a suitably differentiable function of $r$. 
In that case we can write $p_{\T0}$ 
in the following form,
\begin{equation}
p_{\T0}=1+r^2p_{\T2}+r^3p_{\T3}+\cdots
\label{eq:pt0form}
\end{equation} 
where, $p_{\T n}$ is proportional to the $n^{th}$ derivative of the 
initial tangential pressure at the center. We note that choosing the 
form of $\nu_0$ as given by (\ref{eq:nu0form}) automatically fixes 
$p_{\T1}=0$. 
Also using \e (\ref{eq:nu}) in \e (\ref{eq:ein3}), we get,
\begin{equation}
G(t,r)=b(r)e^{2\nu_0(R)}
\label{eq:G}
\end{equation}
Here $b(r)$ is another arbitrary function of $r$. 
In corresponding dust models, we can write, $b(r)=1+f(r)$, 
where $f(r)$ is the velocity distribution function of the 
collapsing shells. In the marginally bound case, $f(r)=0$. We choose
the similar analog here and henceforth consider $b(r)=1$.

The reason for the choice such as above for the mass function, 
and the function $b(r)$, is to bring out the role of pressure 
towards determining the final fate of collapse in a transparent 
manner. For example, in the present situation, if the tangential 
pressure were vanishing identically, the density evolution 
would be necessarily homogeneous throughout, and the collapse will 
necessarily end in a back hole, just as the Oppenheimer-Snyder 
homogeneous dust cloud collapse. On the other hand, a non-vanishing 
pressure gradient gives rise to either one of a naked singularity 
or a black hole as we show below. One can match the cloud to an exterior
by introducing an in between shell, wherein $p_r$ tends to zero at the 
outer boundary of the shell.

Using \e (\ref{eq:G}) in \e (\ref{eq:ein4}), we get,
\begin{equation}
\sqrt{R}\dot{R}=-a(t)e^{2\nu_0(R)}\sqrt{R^3h(R)+\frac{2}{3}r^3-\frac{1}{3}R^3}
\label{eq:collapse}
\end{equation}
Here $a(t)$ is a function of time. By a suitable scaling of the
time coordinate, we can always take $a(t)=1$. The negative sign is 
due to the fact that $\dot{R}<0$ which is the collapse condition. 
The function $h(R)=h(rv)$ is defined as,
\begin{equation}
h(rv)=\frac{e^{2\nu_0(rv)}-1}{r^2v^2}
\label{eq:h}
\end{equation}
Substituting the value of $\nu_0$, the above equation can 
be written as,
\begin{equation}
h(rv)=p_{\T2}+\frac{2}{3}rvp_{\T3}+ \cdots
\label{eq:h1}
\end{equation}
Now simplifying \e (\ref{eq:collapse}), we get,
\begin{equation}
\sqrt{v}\dot{v}=-e^{2\nu_0(rv)}\sqrt{v^3\left(h(rv)-\frac{1}{3}\right)
+\frac{2}{3}}
\label{eq:collapse1}
\end{equation}
Integrating the above equation, we have,
\begin{equation}
t(v,r)=\int_v^1\frac{\sqrt{v}dv}{\sqrt{e^{4\nu_0}\left[v^3\left(h(rv)
-\frac{1}{3}\right)+\frac{2}{3}\right]}}
\label{eq:scurve1}
\end{equation}
We note that the coordinate $r$ is treated as a constant 
in the above equation. Expanding $t(v,r)$ around the center, we get,
\begin{equation} 
t(v,r)=t(v,0)+rX(v)+O(r^2)
\label{eq:scurve2}
\end{equation}
where the function $X(v)$ is given by,
\begin{equation}
X(v)=-\frac{1}{3}\int_v^1dv\frac{v^4\sqrt{v}p_{\T3}}
{\sqrt{v^3\left(p_{\T2}-\frac{1}{3}\right)+\frac{2}{3}}}
\label{eq:tangent1}
\end{equation}
Thus the time taken for the central shell at $r=0$ to reach 
the singularity is given by,
\begin{equation}
t_s(0)=\int_0^1\frac{\sqrt{v}dv}{\sqrt{v^3\left(p_{\T2}
-\frac{1}{3}\right)+\frac{2}{3}}}
\label{eq:scurve3}
\end{equation}
From the above equation it is clear that for $t_s(0)$ to be defined,
$p_{\T2}> 1/3$.
Also, the time taken for the other shells to reach the singularity
can be given as,
\begin{equation}
t_s(r)=t_s(0)+rX(0)+O(r^2)
\label{eq:scurve4}
\end{equation}

To see the dynamic evolution of $p_\T(t,r)$, we put \e (\ref{eq:rho}) 
in \e(\ref{eq:ein2}) and simplify to get,
\begin{equation} 
p_\T(r,v)=1+\frac{1}{2}\left(p_{\T0}(rv)-1\right)\left(\rho(r,v)+1\right)
\label{eq:ptrt}
\end{equation}
It is evident that if at the initial epoch $p_{\T0}=1$, then 
$p_\T(r,v)=1$ throughout the collapse, and the evolution will be
exactly like dust models. In the case otherwise, we find the limiting value 
of $p_{\T0}$ at the central shell as,
\begin{equation}
\lim_{v\rightarrow 0}\lim_{r\rightarrow 0}p_\T(r,v)= 
\lim_{v\rightarrow 0}\lim_{r\rightarrow 0}\left(1+\frac{r^2}{v}\right)
\label{eq:ptlimit}
\end{equation}
It is easily seen that near the central singularity there exist 
ingoing families of {\it timelike} curves of the form,
\begin{equation}
t_{s_0}-t=kr^{\beta}; k>0, \beta>2
\label{eq:family}
\end{equation}
along which the tangential pressure necessarily diverges as we approach
the point $(t_{s_0},0)$ on the $(t,r)$ plane. Hence there always exist 
some timelike paths along which both the tangential pressure and the 
density diverge in the limit of approach to the central singularity.
Thus, as opposed to certain cases, where the pressures are necessarily
bounded at the naked singularity (e.g. in the case of [1]), which is 
somewhat artificial situation, we deal here with a physically more
realistic scenario where pressures diverge at the singularity and then
it is to be seen if a naked singularity is allowed in such a situation.

In order to decide the final fate of collapse in terms of either
a black hole or a naked singularity, we need to study the 
behaviour of the apparent
horizon, and to examine if there are any families of outgoing
nonspacelike trajectories, which terminate in the past at the 
singularity. 
The apparent horizon within the collapsing cloud is given by 
$R/F=1$, which gives the boundary of the trapped surface region 
of the space-time. If the neighborhood of the center gets trapped 
earlier than the singularity, then it will be covered and a black hole
will be the final state of the collapse. In the case otherwise, 
the singularity can be naked with non-spacelike future directed 
trajectories escaping from it to outside observers.

To consider the possibility of existence of such 
families, and to examine the nature of the central singularity 
occurring at $R=0$, $r=0$ in the present case, let us consider the 
outgoing radial null geodesics equation,
\begin{equation}
\frac{dt}{dr}=e^{\psi-\nu}
\label{eq:null1}
\end{equation}
The central singularity occurs at $v=0,r=0$, which 
corresponds to $R=0,r=0$. Therefore, if we have any 
future directed null geodesics terminating in the past at the 
singularity, we must have $R\rightarrow0$ as $t\rightarrow t_s(0)$ 
along the same. 
Now writing the geodesic equation \e (\ref{eq:null1}) in terms 
of the variables $(u=r^{\frac{5}{3}},R)$, we have,
\begin{equation}
\frac{dR}{du}=\frac{3}{5}r^{-\frac{2}{3}}R'\left[1+\frac{\dot{R}}{R'}
e^{\psi-\nu}\right]
\label{eq:null2}
\end{equation}
Using \e (\ref{eq:ein4}) and considering $\dot{R}<0$, we get,
\begin{equation}
\frac{dR}{du}=\frac{3}{5}\left(\frac{R}{u}+\frac{\sqrt{v}v'}
{\sqrt{\frac{R}{u}}}\right)\left(\frac{1-\frac{F}{R}}{\sqrt{G}
(\sqrt{G}+\sqrt{H})}\right)
\label{eq:null3}
\end{equation}
If the radial null geodesics terminate at the singularity in 
the past with a definite tangent, then at the singularity the 
tangent to the geodesic $\frac{dR}{du}>0$, in the $(u,R)$ plane,
with a finite value. In the present case, all singularities for $r>0$ are 
covered since $\frac{F}{R}\rightarrow\infty$ in that case, and hence 
$\frac{dR}{du}\rightarrow-\infty$. Therefore, only the singularity 
at the central shell could be naked. Now from \e (\ref{eq:collapse1}) we 
get for $r\rightarrow 0$ along a constant $v$ line,
\begin{equation}
\sqrt{v}v'=X(v)\sqrt{v^3\left(p_{\T2}-\frac{1}{3}\right)+\frac{2}{3}}
\label{eq:vdash}
\end{equation}
Let us define the tangent to the null geodesics from the 
singularity as,
\begin{equation}
x_0=\lim_{t\rightarrow t_s}\lim_{r\rightarrow 0} \frac{R}{u}
=\left.\frac{dR}{du}\right|_{t\rightarrow t_s;r\rightarrow 0}
\end{equation}
Using \e (\ref{eq:null3}), we get,
\begin{equation}
x_0^{\frac{3}{2}}=\frac{5}{\sqrt{6}}X(0)
\end{equation}
Now, if $p_{\T3}<0$, then $x_0>0$ and hence we would have 
radially outgoing null geodesic coming out from the singularity, 
and the singularity will be naked. While if $p_{\T3}>0$, we will get 
a black hole solution. If $p_{\T3}=0$, then we will have to go 
to the higher order terms and do the same analysis. In the $(t,r)$ plane, 
the equation for the radial null geodesic coming out from the 
singularity is,
\begin{equation}
t-t_s(0)=x_0r^{\frac{5}{3}}
\end{equation}
Also, we see that in case of a naked 
singularity, the singularity curve at the center, \e(\ref{eq:scurve2}), 
is an increasing function of $r$, as in that case $X(v)>0$. 
On the other hand, a black-hole solution gives a decreasing or 
constant curve for shells with increasing $r$. 
It is relevant to note here that $X(0)>0$ implies $v'>0$, and so
from $R'=v+rv'$ we see that there are no shell-crossings, at least 
in a neighbourhood of the central singularity.
We thus see how a non-vanishing pressure causes a naked singularity
as the end state of collapse.

It is also interesting to note how the non-vanishing pressure gradient 
affects the spacetime shear. From \e (\ref{eq:rho}), we get for small
values of $r$ along constant $v$ line,
\begin{equation}
\rho(t,r)=\frac{2}{v^3+rv\sqrt{v}X(v)\sqrt{v^3\left(p_{\T2}
-\frac{1}{3}\right)+\frac{2}{3}}}-1
\label{eq:shear} 
\end{equation}
If $p_{\T3}<0$, then $X(v)>0$, i.e. $v'>0$. Thus from the above 
equation it is evident that $\rho(t,r)$ is a decreasing function of 
$r$ at any given time $t$. But it is known [8] that in the case of 
a vanishing shear, for the mass function we have considered here 
in \e (\ref{eq:mass}), $\rho =\rho(t)$, i.e. the density 
is necessarily homogeneous throughout the collapse. Therefore, 
we conclude that in the case of a naked singularity developing 
as collapse end state, the collapsing cloud has non-zero shear. 
This shear may play the role of deforming the apparent horizon, 
thus exposing the singularity.


\begin{references}


\bibitem{1}
H. Muller zum Hagen, P. Yodzis and H. Seifert, Commun. Math. Phys.
{\bf 37}, 29 (1974).

\bibitem{2}
L. Herrera and N. O. Santos, Physics Reports {\bf 286}, 53 (1997).

\bibitem{3}
P. S. Joshi, {\it Global aspects in gravitation and cosmology},
Clarendon Press, OUP (1993).

\bibitem{4}
M. Celerier and P. Szekeres, gr-qc/0203094;
R. Giambo', F. Giannoni, G. Magli, P. Piccione, gr-qc/0204030.

\bibitem{5}
T. Harada, H. Iguchi, and K. Nakao, Prog.Theor.Phys. {\bf 107}, 449 (2002). 


\bibitem{6}
S. W. Hawking and G. F. R. Ellis, {\it The large scale structure of
spacetime}, Cambridge University Press, Cambridge.


\bibitem{7}
P. S. Joshi and I. H. Dwivedi, Class. Quantum Grav. {\bf 16}, 41 (1999). 


\bibitem{8}
P. S. Joshi, N. Dadhich and R. Maartens, Phys. Rev. D65:
101501, 2002.



\end{references}
\end{document}